\documentclass[11pt] {article}
\usepackage{changebar}
\usepackage{times}
\usepackage{graphicx}
\usepackage{makeidx}
\usepackage{natbib}
\usepackage{latexsym}
\usepackage{bm}
\usepackage{amsmath}
\usepackage{amsthm}
\usepackage{amssymb}
\makeindex
\newcommand{\beq}{\begin{equation}}
\newcommand{\eeq}{\end{equation}}

\newcommand{\defn}{\begin{quote}{\bf Definition. }}
\newcommand{\edefn}{\end{quote}}
\newcommand{\thm}{\begin{theorem}}
\newcommand{\ethm}{\end{theorem}}

\newcommand{\bmat}[1]{\left ( \begin{array}{#1}}
\newcommand{\emat}{\end{array}\right )}
\theoremstyle{definition}

\theoremstyle{plain}
\newtheorem{theorem}{Theorem}

\begin{document}
\vspace*{-3cm}

\begin{center}
{\bf \large COVID-19 and the difficulty of inferring epidemiological parameters from clinical data}

\bigskip

Simon N. Wood\footnote{{\tt simon.wood@bristol.ac.uk} University of Bristol, UK. $^2$ Universit\`a della Svizzera italiana, Switzerland.}, Ernst C. Wit$^2$, Matteo Fasiolo$^1$ and Peter J. Green$^1$. \\ {\small \em Accepted Lancet Infectious Diseases}.

\end{center}

{\small Knowing the infection fatality ratio (IFR) is crucial for epidemic management: for immediate planning; for balancing the life years saved against those lost to the consequences of management; and for evaluating the ethics of paying substantially more to save a life year from the epidemic than from other diseases. Impressively, \cite{verity2020ifr1} rapidly assembled case data and used statistical modelling to infer the IFR for COVID-19. We have attempted an in-depth statistical review of their paper, eschewing statistical nit-picking, but attempting to identify the extent to which the (necessarily compromised) data are more informative about the IFR than the modelling assumptions. First the data.
\begin{itemize}
\item Individual level data for outside China appear problematic, because different countries have differing levels of ascertainment and different disease-severity thresholds even for classification as a case. Their use in IFR estimation would require country-specific model ascertainment parameters, about which we have no information. Consequently these data provide no useful information on IFR.
\item Repatriation flight data provide the sole information on Wuhan prevalence (excepting the lower bound of confirmed cases). 689 foreign nationals eligible for repatriation are doubtfully representative of the susceptible population of Wuhan. Hence it is hard to see how to usefully incorporate the 6 positive cases from this sample. 

\item Case-mortality data from China provide an upper bound for IFR, and, with extra assumptions, on the age dependence of IFR. Since prevalence is unknown, they contain no information for estimating the absolute IFR magnitude.

\item Because of extensive testing, the Diamond Princess (used only for validation by Verity et al.) supplies data on both infections and symptomatic cases, with fewer ascertainment problems. These data appear directly informative about IFR. Against this, the co-morbidity load on the DP is unlikely to fully represent any population of serious interest (perhaps fewer very severe but more milder co-morbidities).     
\end{itemize}

\noindent Secondly, the modelling assumptions: we see two primary problems.
\begin{enumerate}
\item Verity et al. correct the Chinese case data by assuming that ascertainment differences across age groups determine case rate differences. Outside Wuhan they replace observed case data by the cases that would have occurred if each age group had the same per-capita observed case rate as the 50-59 group. They assume complete ascertainment for the 50-59s. These are very strong modelling assumptions that will greatly affect the results: but the published uncertainty bounds reflect no uncertainty about them. In Wuhan, the complete ascertainment assumption is relaxed by introducing a parameter, but one for which the data appear uninformative, so the results will be driven by the assumed uncertainty. 
\item Generically, Bayesian models describe uncertainty both in the data and in prior beliefs about the studied system. Only when data are informative about the targets of modelling can we be sure that prior beliefs play a small role in what the model tells us about the world. In this case the data are especially uninformative: we suspect results are mostly the consequence of what our prior beliefs were. 
\end{enumerate}
Taken together these problems indicate that Verity et al.’s IFRs should be treated very cautiously when planning. While awaiting actual measurements, we would base IFRs on the DP data, with the Chinese case-fatality data informing the dependence of IFR on age: in supplementary material we provide a crude Bayesian model with its IFR estimates by age. Corresponding population IFR estimates and 95\% credible intervals are China: 0.43\% (.23,.65), UK: 0.55\% (.30,.82); India 0.20\% (.11,.30). The strong assumptions required, by this approach too, emphasize the need for improved data. We should replace complex models of inadequate clinical data, with simpler models of epidemiological prevalence data from appropriately designed random sampling using antibody or PCR tests. 

}

%\bibliography{/home/sw283/bibliography/simon}
%\bibliographystyle{chicago}

\pagebreak
\begin{center}
{\bf \large Supplementary material for {\em COVID-19 and the difficulty of inferring epidemiological parameters from clinical data}}

\bigskip

Simon N. Wood\footnote{{\tt simon.wood@bristol.ac.uk} University of Bristol, UK. $^2$ Universit\`a della Svizzera italiana, Switzerland.}, Ernst C. Wit$^2$, Matteo Fasiolo$^1$ and Peter J. Green$^1$. 
\
\end{center}

%\bibliography{/home/sw283/bibliography/simon}
%\bibliographystyle{chicago}

\section{A crude IFR model}

We attempt to construct a model for the Diamond Princess (henceforth DP) data and aggregated data from China, with the intention that the DP data informs the absolute magnitude of the IFR while the China data contributes to the estimation of relative IFR by age class. For the Diamond Princess we lump the 80-89 and 90+ age groups into an 80+ group to match the China data, noting that there are no deaths in the 90+ group. We obtained the age of death of the 12 cases from the Diamond Princess Wikipedia page, checking the news reports on which the information was based. One case has no age reported except that he was an adult. Given that there was no mention of a young victim we have assumed that he was 50 or over. 

We adopt the assumptions of \cite{verity2020ifr} of a constant attack rate with age, and that there is perfect ascertainment in one age class, but assume that this is the 80+ age group for the DP. The assumption seems more tenable for the DP population than for China, given that 4003 PCR tests were administered to the 3711 people on board, with the symptomatic and elderly tending to be tested first. However given that the tests were not administered weekly to all people not yet tested positive, from the start of the outbreak, and that the tests are not 100\% reliable, the assumption is still unlikely to be perfect, which may bias results upwards. Unlike  \cite{verity2020ifr} we do not {\em correct} the case data, but adopt a simple model for under-ascertainment by age, allowing some, but by no means all, of the uncertainty associated with this assumption to be reflected in the intervals reported below. We then model a proportion of the potentially detectable cases as being symptomatic, making a second strong assumption that this rate is constant across age classes. This assumption is made because the data only tell us  that there were 314 symptomatic cases among 706 positive tests but not their ages, so we have no information to further distinguish age specific under-ascertainment and age specific rates of being asymptomatic. We then adopt a simple model for the probability of death with age (quadratic on the logit scale). 

For the China data we necessarily use a different attack rate to the DP, but the same model as the DP to go from infected to symptomatic cases (on the basis that this reflects an intrinsic characteristic of the infectious disease). However we assume that only a proportion of symptomatic cases are detected (at least relative to whatever threshold counted as symptomatic on the DP). Furthermore we are forced to adopt a modified ascertainment model for China, and correct for the difference between this and the DP ascertainment model, within the sub-model for China. We assume the same death rates for symptomatic cases in China, but apply the \cite{verity2020ifr} correction for not-yet-occurred deaths, based on their fitted Gamma model, treating this correction as fixed.  

\section{Technical details of the crude IFR model}

\subsection{Diamond Princess component}
In detail, starting with the Diamond Princess, let $\alpha$ be the infection probability, constant for all age classes, $p^c_i$ the probability of an infection to be detectable
%becoming a potentially detectable case 
in age class $i$, $p^s_i$ the  
probability that a detectable case develops symptoms and $p^d_i$ the probability that a symptomatic case dies. $p^c_i p^s_ip^d_i$ is the IFR for age class $i$. Let $a_i$ denote the lower age boundary of class $i$. The models are (i) for the detectability probability
$$
p^c_i = \gamma_1 + (1-\gamma_1) e^{-(a_i - 80)^2/\exp(\gamma_2)}.
$$  
Note the assumption that all cases in the oldest age class are ascertained; (ii) a constant symptomatic probability model,
$$p^s_i = \phi,$$
and (iii) for the probability that a symptomatic case dies,
$$
\text{logit}(p^d_i) = \beta_1 + \beta_2(a_i-40) + \beta_3(a_i-40)^2.
$$
For a case to be recorded on the DP, the person needed to be attacked by the virus, gotten ill and detected at the right moment. In principle, this means that the number of cases in age class $i$ is distributed as a binom$(p_i^c\alpha,n_i)$, where $p^c_i\alpha$ is the probability of gotten ill \emph{and} detected, and $n_i$ is the number of people in age class $i$ on the DP. However, as only 619 out of the 706 cases have their age recorded, we split the cases into 
$$
C_i \sim \text{binom}(p^c_i \alpha, n_i 619/706) ~~~ \text{and} ~~~ C_i^+ \sim \text{binom}(p^c_i \alpha, n_i(1- 619/706))
$$
where $C_i$ are the observed cases of known age and $C_i^+$ are the additional cases, assumed to follow the same age distribution, but not actually recorded by age. Binomial parameters are rounded appropriately. Letting $S_i$ denote the symptomatics among the cases in age group $i$, we have
$$
S_i \sim \text{binom}(p^s_i,C_i + C_i^+).
$$
The deaths among the symptomatics of known age are distributed as
$$
D_i \sim \text{binom}(p^s_i,S_ih_i)
$$
where $h_i$ is the probability of being of known age on death (this is treated as fixed at 1 for ages less than 50, and 11/12 for 50+ given the one victim on the DP for which no age was recorded, except that he was an adult). For the deaths of unknown age, $D_{na}$, (there is one of these) among the symptomatics of unknown age (an artificial category) $S_{na} = \sum_i (1-h_i)S_i$, we have
$$
D_{na} \sim \text{binom}(p_{na},S_{na})
$$
where the probability of death is $p_{na} = \sum_i (1-h_i)S_ip^d_i/S_{na}$.
Finally the total number of symptomatics is modelled as $S_t \sim N(\sum_i S_i,5^2)$, allowing some limited uncertainty in the symptomatic/asymptomatic classification. 

The {\bf actual available data on the DP} are $S_t$, $D_{na}$ and $\{C_i, n_i, D_i \}_{i=0}^{80}$.

\subsection{The Chinese component}
Moving on to the Chinese data, the assumption is that the patterns with age with respect to detection ($p_i^c$), to being symptomatic ($p_i^s$)  and to death ($p_i^d$) are similar, but the  attack rate $\tilde \alpha$ for China is different.
Let $N_i$ be the population size in age class $i$ and $\tilde S_i$ the symptomatics. Then
$$
\tilde S_i \sim \text{binom}(\tilde \alpha p^c_i p^s_i N_i).
$$ 
Unlike on the DP, only a fraction $\delta_i$ of the symptomatics are tested to become cases,
$$
\tilde C_i  \sim \text{binom}(\delta_i, \tilde S_i)
$$
and the (observed) deaths are then distributed as
$$
\tilde D_i \sim \text{binom}(p^d_i p^y_i / \delta_i, \tilde C_i)
$$
where $p^y_i$ is the average probability of a case in class $i$ having died yet, given they will die --- this was treated as a fixed correction and is computed from the \cite{verity2020ifr} estimated Gamma model of time from onset to death, and the known onset times for the cases. The scaling by $\delta_i$ ensures that $p^d_i$ maintains the same meaning between DP and China. We model $\delta_i$ as $\delta_i = \delta p^{cc}_i/p^c_i$ where $p^{cc}_i$ is an attempt to capture the shape of the actual China detectability with age and is defined as $p^{cc}_i = \exp\{-(a_i-65)^2/e^{\gamma_c}\}$. 

\subsection{Priors and posteriors}
We define the following priors using precision and not variance when defining normal densities:
$$
\alpha \sim U(.1,.9), ~~~ \gamma_1 \sim U(.01,.99), ~~~
\gamma_2 \sim N(7.2,.001),
$$
$$
\phi \sim U(.1,.9), ~~~
\beta_1 \sim N(-3.5,.001),~~~ \beta_2 \sim N(0, .001),~~~ \beta_3 \sim N(0, .001),$$
$$
\tilde \alpha \sim U(10^{-4},.5), ~~~ \delta \sim U(.1,.9), ~~~ \gamma_c \sim N(7.4,.01).
$$

This structure uses the information from the DP to assess the symptomatic rate and hidden case rate and the scale of the death probabilities, while the China data refines the information on how death rates change with age. It is possible to formulate a model in which the China data appear to contribute to inference about absolute levels of mortality, but this model is completely driven by the prior put on proportion of cases observed (about which the China data are completely uninformative).    

\begin{figure}[tb]
%	\eps{-90}{.5}{DP-death-pp.eps}
\centerline{
	\includegraphics[scale=.5,angle=-90]{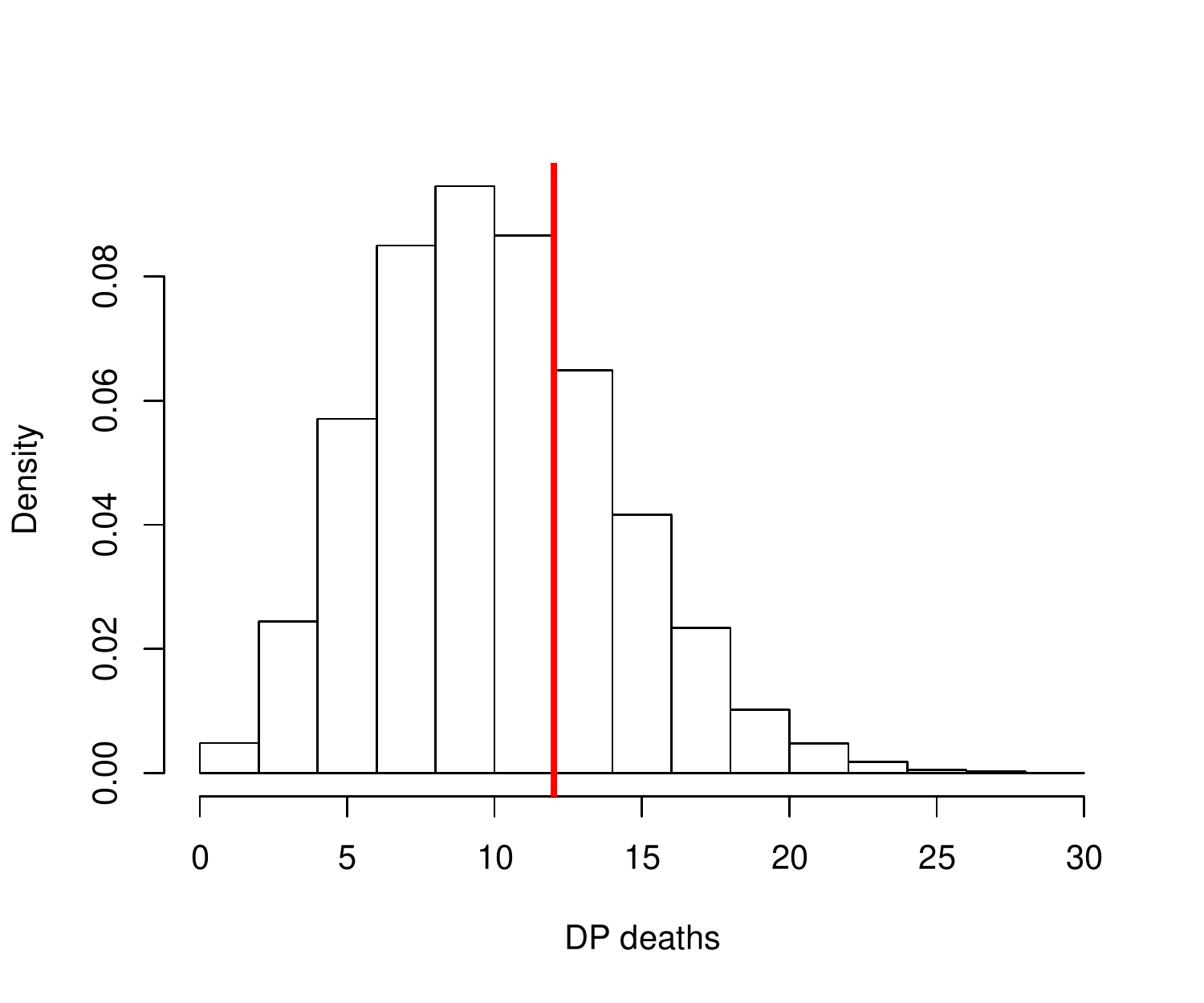}}
	\caption{Posterior predictive distribution of the number of deaths on the Diamond Princess. The vertical red line is the actual number of deaths. \label{fig:DP}} 
\end{figure}

\begin{figure}[tb!]
\includegraphics[scale=.5]{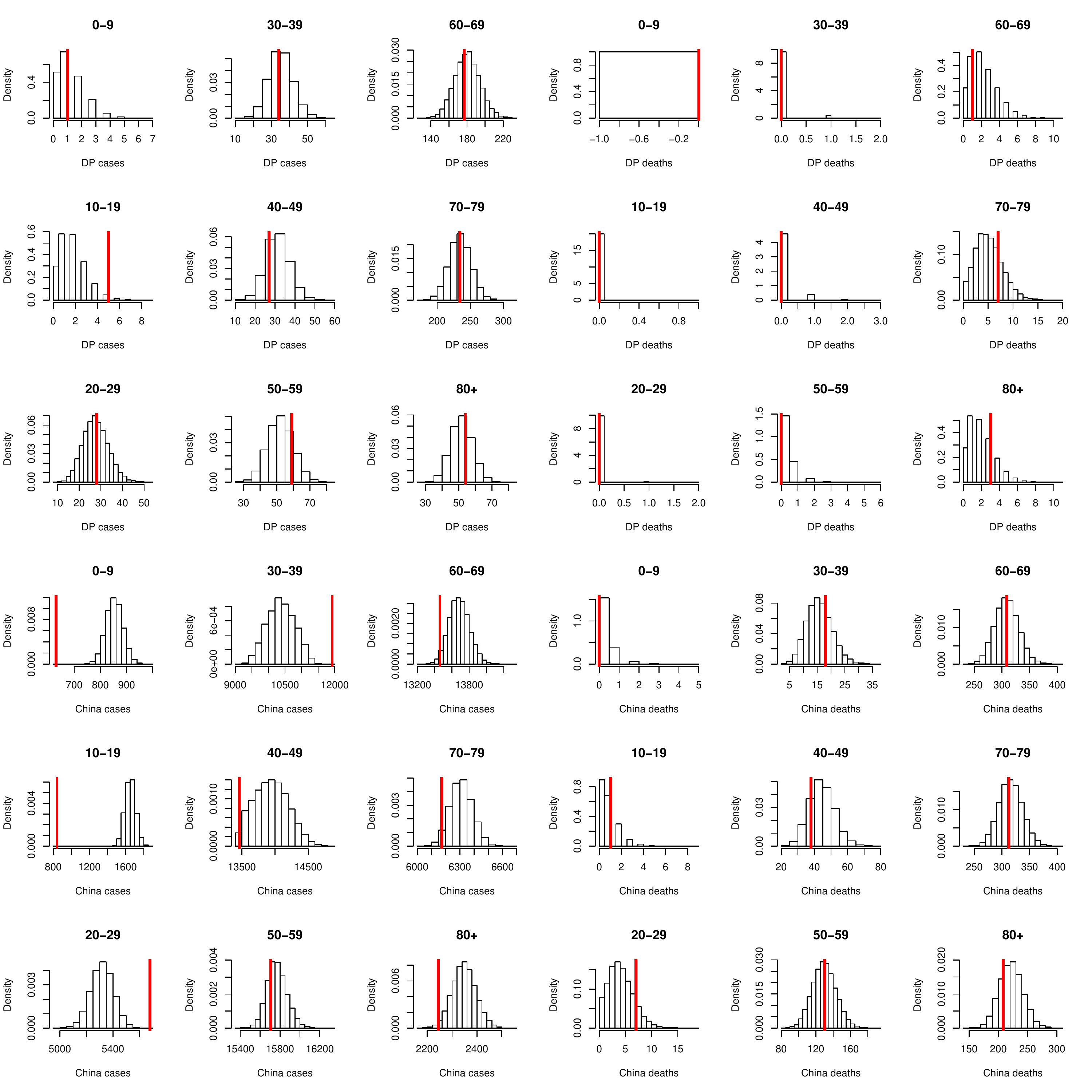}
\caption{The posterior predictive distributions for the cases (left 3 by 3 block) and deaths (right 3 by 3 block) for the DP (top 3 rows) and China (bottom 3 rows) as histograms, with the observed values as vertical red lines. Clearly there are problems with the younger Chinese cases. \label{postpred}} 
\end{figure}

The model was implemented in JAGS 4.3.0. Mixing is slow, but $5\times10^7$ steps, retaining every 2500th sample, gives an effective sample size of about 660 for $\delta$, the slowest moving parameter. We discarded the first 2000 retained samples as burn in, although diagnostic plots show no sign that this is necessary. Posterior predictive distribution plots are shown in Figure~\ref{postpred}. We note the problems with young Chinese detected cases, although even the most extreme mismatch only corresponds to a factor of 2 IFR change, if reflecting incorrect numbers of actual cases. In older groups the model cases are a little high on average, but not by enough to suggest much change in IFR. These mismatches might be reduced by better models for the ascertainment proportion by age. Figure~\ref{fig:DP} shows the posterior predictive distribution for total Diamond Princess deaths with the actual deaths as a thick red bar.

\begin{table}[tb!]
\begin{center}
\begin{tabular}{lll}
Group & median IFR & 95\% Interval\\
\hline
Overall China & 0.43 &(0.23,0.65) \\
Overall UK & 0.55 & (0.30,0.82) \\
Overall India & 0.20 & (0.11,0.30)\\
0-9 & 0.0007 & (0.0002,0.002)\\
10-19 & 0.003 & (0.001,0.006)\\
20-29 & 0.01 & (0.005,0.02)\\
30-39 & 0.03 & (0.02,0.05)\\
40-49 & 0.1 & (0.05,0.15)\\
50-59 & 0.32 & (0.17, 0.50)\\
60-69 & 1.0 & (0.55,1.53)\\
70-79 & 2.3 & (1.2,3.4)\\
80+ & 3.7 & (2.0,5.7)
\end{tabular}
\end{center}
\caption{Posterior median and credible intervals of Infection Fatality Ratio for various groups. We believe the credible intervals to be optimistically narrow.} \label{tab:post} 
\end{table}

The median and credible intervals for the IFR {\em as percentages} in various groups are in Table~\ref{tab:post}. They show different estimates of this crucial quantity compared to \cite{verity2020ifr}, again emphasising the urgent need for statistically principled sampling data to directly measure prevalence, instead of having to rely on complex models of problematic data with strong built in assumptions.

\bigskip

\noindent {\bf Acknowledgements:} we thank Jonathan Rougier and Guy Nason for helpful discussion of onset-to-death interval estimation and the individual level data.

\vfill

\pagebreak

\subsection{JAGS Code}

{\scriptsize\begin{verbatim}
## JAGS code for Diamond Princess + China model
## This version uses Wuhan type case correction for DP also (short
## cleared infections won't show up in DP PCR testing)
## n[i], C[i], St, D[i], DNA and pa[i] are observed nodes
## alternative priors (over informative on prob scale) commented
## out.
model {
  #gamma[1] <- ilogit(lgamma[1]) ## baseline detection prob
  ## probability of becoming a case...
  #alpha <- ilogit(lalpha)
  ## probability of being detected as a case before clearing infection
  for (i in 1:8) {
    pc[i] <- gamma[1] + (1-gamma[1])*exp(-(age[i]-70)^2/exp(gamma[2]))
  }
  pc[9] <- 1
  for (i in 1:9) { ## there are 9 age classes (11 90+ added to 80+)
    ## cases by age are only observed for 619/706, so have to scale prob...
    ## This approach slightly underconstrains model as I know total should
    ## be 706 and have failed to include fact
    Cl[i] ~ dbinom(pc[i]*alpha,n[i]-round(n[i]*619/706)) ## extra cases not classified by age in data
    C[i] ~ dbinom(pc[i]*alpha,round(n[i]*619/706)) ## Observed nodes
    Cpp[i] ~ dbinom(pc[i]*alpha,round(n[i]*619/706)) ## Posterior predictive version
    ## probability of developing symptoms as a case...
    ps[i] <- phi #ilogit(lphi)
    ## the symptomatic....
    S[i] ~ dbinom(ps[i],C[i]+Cl[i])
    Spp[i] ~ dbinom(ps[i],Cpp[i]+Cl[i]) ## Posterior predictive version
    ## probability of death given symptoms and age known...
    pd[i] <- ilogit(beta[1] + beta[2]*(age[i]-40)+beta[3]*(age[i]-40)^2)
    ## deaths in Symptomatics of known age...
    D[i] ~ dbinom(pd[i],round(S[i]*pa[i])) ## Observed node
    Dpp[i] ~ dbinom(pd[i],round(Spp[i]*pa[i])) ## Posterior predictive version
  }
  ## total with symptoms... 
  St <- sum(S) ## Monitor this as posterior predictive
  ## allow some slop in symptomatic/asymptomatic assessment...
  Sy ~ dnorm(St,1/25) ## Observed node
  ## deal with Deaths without an age...
  SNA <- sum((1-pa)*S) ## Symptomatic with unknown age
  DNA ~ dbinom(sum((1-pa)*pd*S)/SNA,round(SNA)) ## Observed node
  for (i in 1:9) {
    pcc[i] <- exp(-(age[i]-65)^2/exp(gamma.ch))
    dc[i] <- pcc[i]/pc[i] ## China detection profile correction 
  }
  ## Now do China, pop is exposed pop...
  #alpha.ch <- ilogit(lalpha.ch) ## China attack rate (assumed constant with age as in paper)
  for (i in 1:9) { ## China data has 9 age classes
    Sch[i]  ~ dbinom(pc[i]*alpha.ch*ps[i],pop[i]) ## potentially detectable symptomatics
    ## but only some proportion of those are detected...
    Cch[i] ~ dbinom(delta*dc[i],Sch[i]) ## observed node
    Cchpp[i] ~ dbinom(delta*dc[i],Sch[i]) ## Posterior predictive version
    Dch[i] ~ dbinom(pd[i]*pyet[i]/(delta*dc[i]),round(Cch[i])) ## observed node
    Dchpp[i] ~ dbinom(pd[i]*pyet[i]/(delta*dc[i]),round(Cchpp[i])) ## Posterior predictive version
  }  
  ## priors...
  #lalpha ~ dnorm(-1.2,0.001) ## logit infection prob
  alpha ~ dunif(.1,.9)
  #lgamma[1] ~ dnorm(-1,.001) ## baseline detection
  gamma[1] ~ dunif(.01,.99)
  gamma[2] ~ dnorm(7.2,.01) ## detection decline rate param
  #lphi ~ dnorm(0,.001)
  phi ~ dunif(.1,.9)
  beta[1] ~ dnorm(-3.5,.001)
  beta[2] ~ dnorm(0,.001)
  beta[3] ~ dnorm(0,.001)
  ## China only
  #lalpha.ch ~ dnorm(-3,.25) ## logit China attack rate
  alpha.ch ~ dunif(1e-4,.5)
  delta ~ dunif(0.1,.9) ##probability of detecion in China given potentially detectable
  gamma.ch ~  dnorm(7.4,.01) ## detection decline rate param China
}
\end{verbatim}}
\subsection{R Code}
{\scriptsize\begin{verbatim}
## Diamond Princess and China model - the two data sources that appear to
## contain information. 
library(rjags)
load.module("bugs")
load.module("glm")

## DP Data for JAGS model
## lower (age) limit in 10 year classes (n)umber in each age class
## (C)ases in each age class, (Sy)mptomatic total
## DNA is Deaths No Age, pa is probability of not knowing age.
## Given the reports it seems reasonable to assume that the
## one case without an age was adult (certain) and over 50
## as there was no reporting of youngest victim etc.
dat <- list(age = 0:8*10 ,
            n = c(16,23,347,429,333,398,924,1015,226), ## DP pop by age class
            C =c(1,5,28,34,27,59,177,234,54), ## cases
	    Sy = 706-392, ## symptomatic cases,
	    DNA=1, ## death of unknown age
            D=c(0,0,0,0,0,0,1,7,3), ## deaths
	    pa = c(rep(1,5),rep(0.92,4))) ## fixed prob death was of known age

## Adding in the China data, aggregated Wuhan and outside...

dat$pop <- c(1273576,1160864,1682459,1659489,1869228,1515041,1157168,533544,229632)
dat$Cch <- c(631,841,5679,11920,13462,15706,13462,6170,2244)
## corrections for insufficient time to see all deaths...
dat$pyet <- c(0.410218110039586,0.410897658916389,0.409538561162789,0.408223434218869,0.407967104385195,
  0.406929578867941,0.404495034342807,0.404179391611681,0.403380149466811)
dat$Dch <- c(0,1,7,18,38,130,309,312,208)

###################
setwd("foo/bar") ## NOTE: set to jags file location
###################

jdp <- jags.model("dp-china.jags",data=dat,n.adapt=10000) ## complie JAGS model

## Sample from JAGS model...

system.time(um <- jags.samples(jdp,
c("pc","ps","pd","delta","alpha.ch","alpha","phi","Cpp","Dpp","St","Cchpp","Dchpp"),
n.iter=50000000,thin=2500)) 

effectiveSize(as.mcmc.list(um$delta))
hist(um$delta)

## look at posterior predictive plots...
ps <- FALSE
if (ps) pdf("post-pred.pdf",height=12,width=12)
main <- c("0-9","10-19","20-29","30-39","40-49","50-59","60-69","70-79","80+")
lay <- matrix(0,6,6)
lay[1:3,1:3] <- 1:9; lay[1:3,4:6] <- 1:9 + 9
lay[4:6,1:3] <- 1:9 + 2*9;lay[4:6,4:6] <- 1:9 + 3*9
layout(lay)
drop <- 1:2000 ## burn-in
for (k in  1:4) {
  if (k==1) { pp <- um$Cpp;true <- dat$C;xlab <- "DP cases"} else
  if (k==2) { pp <- um$Dpp;true <- dat$D;xlab <- "DP deaths"} else
  if (k==3) { pp <- um$Cchpp;true <- dat$Cch;xlab <- "China cases"} else
  { pp <- um$Dchpp;true <- dat$Dch;xlab <- "China deaths"}
  for (i in 1:9) {
    hist(c(pp[i,-drop,1],true[i]),main=main[i],xlab=xlab,freq=FALSE) ## posterior predictive
    abline(v=true[i],lwd=3,col=2) ## truth
  }
}
if (ps) dev.off()

## IFR histograms and credible intervals...

par(mfrow=c(3,3))
ci <- matrix(0,3,9)
mode.p <- mean.p <- rep(0,9)
for (i in 1:9) {
  ifr <- um$ps[i,,1]*um$pd[i,,1]*um$pc[i,,1]
  ifr <- ifr[-(1:2000)]
  x <- hist(log10(ifr),
            main=(i-1)*10,xlab="log10(risk)")
  ci[,i] <- quantile(ifr,c(.025,.5,.975))
  mode.p[i] <- 10^x$mid[x$counts==max(x$counts)]
  mean.p[i] <- mean(ifr)
}
ci*100
mode.p*100

## sanity check against DP deaths...
if (ps) postscript("DP-death-pp.eps",width=6,height=5)
hist(colSums(um$Dpp[,-(1:2000),1]),xlab = "DP deaths",main="",freq=FALSE);abline(v=12,lwd=3,col=2)
if (ps) dev.off()

## various demographies...

demog <- c(.1,.1,.15,.15,.17,.14,.1,.05,.04) ## roughly China demography
## Wikipedia Indian demography...
india <- c(.198,0.2091,0.1758,0.1435,0.1113,0.0728,0.0529,0.0235,.0131)
## https://www.statista.com/statistics/281174/uk-population-by-age/
uk <- c(8.05,7.53,8.31,8.83,8.5,8.96,7.07,5.49,3.27) ## total pop statista
uk <- uk/sum(uk) ## 2018 UK demography

## Verity point estimate IFR by age...
verity <- c(.000161,.00695,.0309,.0844,.161,.595,1.93,4.28,7.80)/100 

sum(dat$C*verity) ## DP deaths according to Verity and assuming all cases found
sum(uk*verity)  ## UK IFR Verity point estimates
sum(uk*ci[2,])  ## UK IFR median point estimates

## overall IFR for various demographies...

pt <- demog %*% (um$ps[,,1]*um$pd[,,1]*um$pc[,,1]) 
quantile(pt,c(.025,.5,.975))*100 ## China
pt <- uk %*% (um$ps[,,1]*um$pd[,,1]*um$pc[,,1]) 
quantile(pt,c(.025,.5,.975))*100 ## UK
pt <- india %*% (um$ps[,,1]*um$pd[,,1]*um$pc[,,1]) 
quantile(pt,c(.025,.5,.975))*100 ## India
\end{verbatim}}

\end{document}